\documentclass[times]{aa}
\usepackage{psfig}
\usepackage{natbib}

\bibpunct{(}{)}{;}{a}{}{,}

\begin{document}

\def\eps{\varepsilon}
\def\aap{A\&A}
\def\apj{ApJ}
\def\apjl{ApJL}
\def\mnras{MNRAS}
\def\aj{AJ}
\def\nat{Nature}
\def\aaps{A\&A Supp.}
\def\degs{$^\circ $}
\def\msun{{\,M_\odot}}
\def\rsun{{\,R_\odot}}
\def\lsun{{\,L_\odot}}
\def\sgra{Sgr~A$^*$}
\def\medd{\dot{M}_{\rm Edd}}
\def\ledd{{L}_{\rm Edd}}
\newcommand\mdot{\dot{m}}
\def\simlt{\lower.5ex\hbox{$\; \buildrel < \over \sim \;$}}
\def\simgt{\lower.5ex\hbox{$\; \buildrel > \over \sim \;$}}

\def\del#1{{}}
\def\C#1{#1}

\title{Self-gravitating accretion disk in Sgr A* few million years
ago: was Sgr A* a failed quasar?}
\titlerunning{Sgr A* few Million years ago}
\author{Sergei Nayakshin \and  Jorge Cuadra} 
\authorrunning{S. Nayakshin \and J. Cuadra}
\institute{Max-Planck-Institut f\"{u}r Astrophysik,
Karl-Schwarzschild-Str.1, D-85741 Garching, Germany
}
\date{Submitted to A\&A. September 22, 2004} 
\offprints{S. Nayakshin,\\ e-mail: \tt{serg@MPA-Garching.MPG.DE}}

\abstract{Sgr A* is extra-ordinarily dim in all wavelengths requiring
a very low accretion rate at the present time. However, at a radial
distance of a fraction of a parsec from Sgr A*, two rings populated by
young massive stars suggest a recent burst of star formation in a
rather hostile environment. Here we explore two ways of creating such
young stellar rings with a gaseous accretion disk: by self-gravity in
a massive disk, and by capturing ``old'' low mass stars and growing
them via gas accretion in a disk. The minimum disk mass is above $10^4
\msun$ for the first mechanism and is few tens times larger for the
second one. The observed relatively small velocity dispersion of the
stars rules out disks more massive than around $10^5 \msun$: heavier
stellar or gas disks would warp each other by orbital precession in an
axisymmetric potential too strongly. The capture of ``old'' stars by a
disk is thus unlikely as the origin of the young stellar disks. The
absence of a massive nuclear gas disk in \sgra\ now implies that the
disk was either accreted by the SMBH, which would then imply almost a
quasar-like luminosity for \sgra, or was consumed in the star
formation episode.  The latter possibility appears to be more likely
on theoretical grounds. We also consider whether accretion disk plane
changes, expected to occur due to fluctuations in the angular momentum
of gas infalling into the central parsec of a galaxy, would dislodge
the embedded stars from the disk midplane. We find that the stars
leave the disk midplane only if the disk orientation changes on time
scales much shorter than the disk viscous time.}  \maketitle

\section{Introduction}

The complex chain of events leading to the growth of supermassive
black holes (SMBHs) in the galactic centers is not yet fully
understood \citep[e.g.,][]{Rees02}. Nevertheless, gas accretion is
probably the dominant physical process delivering the food
to the giant black holes \citep[e.g.,][]{Yu02}. On small radial
distances from the SMBH, the standard thin accretion disk
\citep{Shakura73} appears to be a sure way to provide the SMBH with
gas at rates approaching the Eddington limit.  However, at distances
larger than $\sim 10^{-2}$ parsec from the SMBH, standard thin disks
run into several problems.

First of all, the time necessary for the gas to inflow into the black hole --
the disk viscous time scale -- becomes too long (e.g. as large as
$10^8-10^{10}$ years for larger radii). In parallel with this, the (standard
accretion) disk mass becomes very large. When the latter exceeds about 1 \% of
the SMBH mass, local gravitational instabilities develop
\citep[e.g.,][]{Paczynski78, Kolykhalov80, Shlosman89, Goodman03,
Collin99}. The structure of such disks is very much open to discussion (see \S
\ref{sec:discussion}) due to uncertainties in theory and a dearth of relevant
observations.

\sgra\ is the closest SMBH \citep[with a mass $M_{\rm BH} \approx 3\times
10^6\msun$; e.g.,][]{Schoedel02}.  Although \sgra\ remained extremely dim
during the entire history of X-ray observations, there are hints that it was
much more active in the past. X-ray and $\gamma$-ray spectrum of the giant
molecular cloud Sgr B2 is most naturally explained as a time-delayed
reflection of a source with a flat AGN-like spectrum
\citep[e.g.,][]{Sunyaev93, Koyama96, Revnivtsev04}. The required luminosity is
in the range of $\sim$few~$\times 10^{39}$ erg/sec, too high by the Galactic
standards. \sgra is then strongly suspected of being brighter in X-rays by
some 6 orders of magnitude 300-400 years ago. 

Deeper in the past, some few million years ago, few dozens of massive stars
were formed and are currently at a distance of order $0.1-0.3$ pc from \sgra\
\citep{Krabbe95, Genzel03, Ghez03b}. This is surprising given that the tidal
force of the SMBH would easily shear even gas clouds with densities orders of
magnitude higher than the highest density cores of GMCs observed in the
Galaxy.

{\em In situ} star formation scenarios for the \sgra\ young massive
stars have been numerically studied by \cite{Sanders98} with a
sticky-particle code and also qualitatively described by
\cite*{Morris99}. In particular, one of the simulations done by
\cite{Sanders98} assumed that a cold cloud of gas with radial dimensions
of $0.4$ pc and with a small angular momentum infalls into \sgra\
gravitational potential starting from distance of 2.4 parsec. The
cloud gets tidally sheared into a thin band of gas which then forms a
precessing eccentric ring. Frequent shocks are assumed to lead to
strong compression of the gas and star formation.

The existence of such low angular momentum clouds seems to be in
question. In addition, the initial conditions of the simulations are
rather extreme: to be stable against tidal shear
\cite[e.g., eq. 1 of][]{Sanders98}, the cloud mass should be
$M_{cl}\simgt 7\times 10^4 \msun$, a very large mass for a cloud of
0.4 pc in size.  The recent observations \citep{Liszt03} seem to
contradict the \cite{Sanders98} model for the ionized gas streamers.
\cite{Genzel03} discount a current star formation in the mini-spiral,
which is believed to be an ionized streamer. \cite{Genzel03} also note
more generally that ``if massive stars are forming frequently in dense
gas streamers when outside the central parsec and then rapidly move
through the central region, one would expect $\sim 100$ times as many
massive stars outside the central region as in the central parsec'',
which is not the case observationally.

Alternatively to the {\em in situ} star formation, \cite{Gerhard01}
proposed that the young massive stars could have been formed outside
the central parsec in a massive star cluster. Then, due to dynamical
friction with the older population of background stars, the cluster
would have been dragged into the central parsec and then dissolved
there by the SMBH tidal shear.  However, this appears to be only
possible \citep{Kim03, McMillan03} if the cluster is very massive
($M\simgt 10^6\msun$), or if it is formed very near the central parsec
already. In both cases a very dense core for the star cluster is
required and appears to be unrealistic. An intermediate mass black
hole in the center of the cluster does allow the star cluster to
survive longer against tidal disruption and hence transport the young
stars in the central parsec more efficiently \citep[as suggested
by][]{Hansen03}. However the numerical simulations of \cite*{Kim04}
show that the mass of the black hole has to be unusually large ($\sim
10$\%) compared with the cluster mass for this idea to work in
practice.

\cite{Levin03} and Nayakshin, Cuadra, \& Sunyaev (2004) suggested that
the origin of the young stars is a massive self-gravitating accretion
disk existing in \sgra\ in the past. Here we intend to investigate
this idea quantitatively and to also look into some related
theoretical questions.

We first estimate the minimum mass of such accretion disks to be
around $10^4$, for each of the two stellar rings. In addition, we rule
out the possibility that a less massive accretion disk could capture
enough of low mass stars from the pre-existing ``relaxed'' \sgra\ cusp
and then grow them by accretion into massive stars (\S
\ref{sec:capture}).

We then attempt to understand the spatial distribution of the young
stars. In particular, we find that the rate of {\it N}-body scattering
between the stars (\S \ref{sec:group}) of the same ring can explain
the observed stellar velocity dispersions in the inner stellar ring if
the time-averaged total stellar mass in the ring was $10^4 \msun$ or
higher. We also find that stellar orbits in both rings should remain
close to the circular Keplerian orbits up to this day (if stars were
indeed born in a disk).  The outer ring is however observed to be
geometrically thicker and with a higher velocity dispersion than the
inner one. The velocity dispersion of the outer ring may result from
the stellar disk warping in the gravitational potential of the inner
ring. Such warping sets the upper limit on the disk mass of about
$10^5 \msun$ (\S \ref{sec:max}).

We also question in \S \ref{sec:two} whether it is possible for the
disk to leave the newly born stars behind (due to their high inertia)
when the disk plane rotates. Dislodging the newly born stars, or
proto-stars, from the disk midplane would have significantly reduced
the problems faced by accretion disks at large radii since these stars
would then stop devouring the disk and instead heat it and speed up
the accretion of gas onto the SMBH via star-disk collisions
\citep{Ostriker83}.  However, we find that the disk maintains a firm
grip on these stars unless the plane change occurs on a time scale
much shorter than the disk viscous time (\S \ref{sec:two}).

It is found that young massive stars would not migrate much radially
(\S \ref{sec:accretion}) in the disk, meaning that they are probably
located at the radius where they were originally formed. Small scale
proto-stellar disks around the embedded stars may be gravitationally
unstable as well and may create further generations of
stars. Hierarchical growth and merging of such objects may result in
the creation of ``mini star clusters'' with the central object
collapsing to an intermediate mass black hole (\S
\ref{sec:cluster}). This could potentially be relevant to the
observations of such objects as IRS13 \citep{Maillard04}.

Since the combined mass of the stellar material in the observed
stellar rings presently is $\simlt 10^3 \msun$ \citep{Genzel03}, there
is then an interesting question of whether most of the gaseous disk
mass has been used to activate the presently dormant \sgra\ or it was
reprocessed through star formation and expelled to larger radii via
winds and supernova explosions. We believe the latter outcome is more
likely since the accretion of gas onto embedded stars is very
efficient. We briefly discuss observations that could distinguish
between the quasar and the nuclear starburst possibilities (\S
\ref{sec:discussion}).

\del{This star forming activity should have required at least
thousands of $\msun$ of gas.  unlikely to have gone without activation
of the SMBH since the latter would clearly have captured a good
fraction of the gas unused in star formation.}

\section{The minimum mass of a self-gravitating disk in \sgra\ is 10$^4
\msun$}\label{sec:selfgrav}

The standard accretion disk solution \citep{Shakura73} neglects
self-gravity of the disk. Clearly, this solution becomes invalid when
the disk becomes strongly self-gravitating, but here we only want to
estimate the minimum disk mass at which the self-gravity becomes
important. For numerical values of the standard disk parameters, we
follow \cite{Svensson94} with their parameters $\xi=1$ and $f=0$
(i.e. no X-ray emitting disk corona is assumed). The dimensionless
accretion rate $\dot m$ is defined as $\dot m = \dot{M}/\medd$, where
$\medd = \ledd/\varepsilon c^2$ is the accretion rate corresponding to
the Eddington luminosity and $\varepsilon \approx 0.06$ is the
radiative efficiency of the standard accretion flow. \footnote{Note
that our definition of $\mdot$ is different from the one
\cite{Svensson94} use: $\dot m_{\rm SZ} = 17.5 \dot m\,$.}

\begin{figure}
\centerline{\psfig{file=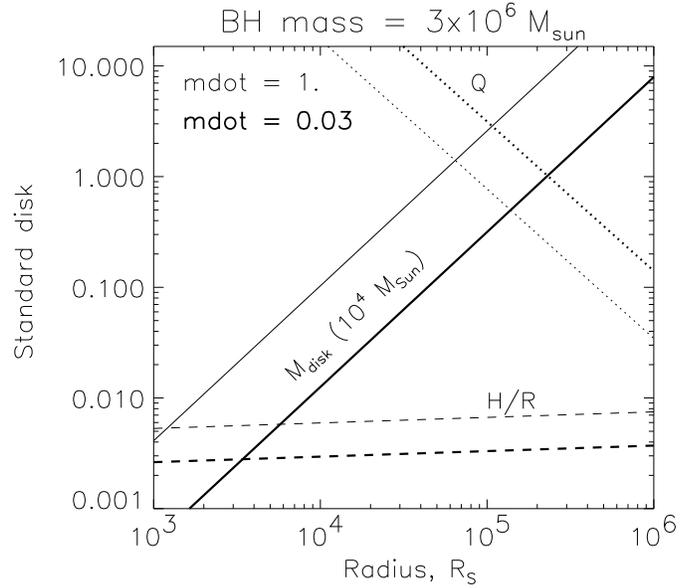,width=.49\textwidth}}
\caption{The disk mass in units of $10^4 \msun$, the Toomre
instability parameter $Q$, and the ratio of the disk height scale $H$
to radius $R$ as a function of radius for the standard accretion disk
model. The thick curves are plotted for $\mdot=0.03$, whereas thin
ones are for $\mdot=1$. In both cases the disk is unstable to
self-gravity at $R\simgt 10^5 R_{\rm S}$, where its mass is $M_{\rm d} \simgt 10^4
\msun$.}
\label{fig:disk3e6a}
\end{figure}

For large radii ($r \gg 1$) the gas dominated equations are appropriate:
\begin{eqnarray}
\frac{H}{R} = 2.7 \times10^{-3} \left(\alpha M_8\right)^{-1/10} r^{1/20} 
\dot m^{1/5} \; , \\
\Sigma = 4.2 \; \times10^6\; \hbox{g cm}^{-2} \; \alpha^{-4/5}\;
M_8^{1/5} r^{-3/5} \dot m^{3/5} \; . \label{sigma}
\\
T = 6.3 \times10^2\; \hbox{K}\; (\alpha M_8)^{-1/5} 
\dot m^{2/5} \left[\frac{R}{10^5 R_S}\right]^{-9/10} \;.
\end{eqnarray} 
Where $H$ is the disk vertical height scale, $R$ is the distance from
the SMBH, $T$ is the midplane gas temperature, $\alpha$ is the
dimensionless viscosity parameter, $M_8 = M_{\rm BH} / 10^8\msun$, $r
= R/R_{\rm S}$, $R_S = 2GM_{\rm BH}/c^2$ is the Schwarzschild
radius of the SMBH and $\Sigma$ is the surface density of the
accretion disk.
These equations assume Thomson electron scattering opacity for
simplicity. Figure \ref{fig:disk3e6a} shows some of the disk
parameters obtained for $M_{\rm BH} = 3\times 10^6 \msun$, $\alpha =
1$, and two values of the dimensionless accretion rate, $\mdot = 0.03$
(thick lines) and $\mdot = 1$ (thin lines). The mass of the disk as a
function of radius is
approximated as $M_{\rm d} \approx \pi R^2 \Sigma$.

It is well known that the standard accretion disk becomes self-gravitating at
large radii, when the \cite{Toomre64} parameter $Q$ becomes less than unity,
\begin{equation}
Q = \frac{c_{\rm s} \Omega}{\pi G \Sigma} \approx \frac{H}{R}
\frac{M_{\rm BH}}{M_{\rm d}}< 1 \;
\label{qdef}
\end{equation}
($c_{\rm s}$ is the sound speed inside the disk and $\Omega$ its
angular velocity). The radius where $Q=1$ yields the minimum mass of
the disk needed for the latter to become self-gravitating. As can be
seen from Fig. \ref{fig:disk3e6a}, the disk should weigh at least
$10^4 \msun$ in order to become self-gravitating. Note that this
minimum disk mass estimate is quite robust because $H/R$ depends on
$\alpha$, radius and the accretion rate only weakly.

This estimate is also conservative. The basic Shakura-Sunyaev model
used here does not include irradiation by the central source, which
may increase the disk midplane temperature somewhat, leading to a
slightly larger $H/R$. In addition, trapping radiation by opacity
effects would reduce the efficiency of cooling, adding to the
stability of the disk against self-gravity. 

\del{A better estimate of the minimum mass would include a more careful
energy balance for the disk, e.g. taking into account the dust
opacity. }

\section{Capturing low mass stars and growing them by
accretion: too slow.}\label{sec:capture} 

\cite*{Artymowicz93} noted that stars in the nuclear star cluster on orbits
relatively close to the local circular rotation of the accretion disk in
quasars will be captured by the disk. The stars can then rapidly grow by gas
accretion, and then enrich the accretion disk with heavy elements through
stellar evolution. For the problem of the observed young massive stars in the
Galactic Center (GC), the Artymowicz et al. trapping mechanism may be an
alternative route to form the stars. The accretion disk does not have to be
self-gravitating for the mechanism to work, provided there is enough stars and
the time scale for the star trapping is right.  One may thus hope to reduce
the required disk mass.

To within a factor order unity, equation 15 of \cite{Artymowicz93}
yields the number of stars  captured by
the disk within time $\Delta t$ as
\begin{equation}
dN_*(R) \sim \frac{\zeta^4}{4} N_*(R) \;,
\label{art}
\end{equation}
where $N_*$ is the total number of stars in the star cluster within
radius $R$, and the variable $\zeta$ is defined by
\begin{equation}
\zeta^4 = 32 C_{\rm d} \frac{M_* M_{\rm d}}{M_{\rm BH}^2} \frac{\Delta t}{P}\;,
\end{equation}
where $P$ is the orbital period and $M_* = m_* \msun$ is the typical
mass of the stars in the cluster. For an estimate, we take $C_{\rm
d}\simeq 3$ and $q_{-2} \equiv 100 M_{\rm d}/M_{\rm BH} \simlt 1$.
Note also that we want to start with abundant stellar seeds, so we
assume $m_*\sim 1$.  At the typical radial position of the young
massive stars in \sgra, $R=0.1 \hbox{pc} \sim 3\times 10^5 R_{\rm S}$,
the circular Keplerian rotation period is
\begin{equation}
P\simeq 10^3 \;\hbox{year}\; \left[\frac{R}{3\times 10^5 R_S}\right]^{3/2}
\label{period}
\end{equation}
Thus,
\begin{equation}
\zeta^4\simeq 3\times 10^{-4} m_* q_{-2} \frac{\Delta t}{10^3 P}\;.
\label{zeta}
\end{equation}
Now, from results of \cite{Genzel03} we estimate that 
\begin{equation}
N_*(R) \simeq 10^5 \; m_*^{-1} \, \left[\frac{R}{3\times 10^5
R_S}\right]^{3/2}
\label{nstar}
\end{equation}
\citep[we assumed the cusp power-law index $p=1.5$ for simplicity; see also
equation 15 in][]{NCS04}. Therefore, the number of stars
captured by the disk is
\begin{equation}
dN_* \sim \; 7 \; q_{-2} \, \frac{\Delta t}{10^6 \; \hbox{years}}\;.
\label{dN}
\end{equation}
Note that the radius $R$ and the average stellar mass $m_*$ scaled out
of this relation.

The number of captured stars is somewhat low if we take into account
the fact that the $\sim 12$ ``Helium'' stars found in each of the
rings are only the brightest end of the stellar distribution, and there
are probably many more (less massive) stars in the rings (\S 3.7 in
\cite{Genzel03}). Therefore one would require $q_{-2} \simgt 3-10$,
that is disk mass $M_d \sim (1-3)\times 10^5 \msun$. With this rather
high required disk mass, the disk would have to be self-gravitating,
and one expects a large number of stars being born inside the
disk. The disk capture mechanism thus fails to reduce the minimum disk
mass.  Nevertheless, it should not be forgotten completely because of
its ability to bring in some {\em late} type stars into the plane of
the disk, e.g. the plane of the young massive stars. This type of
stars would not be born inside the accretion disk in a time span of
just a few million years.

\section{Velocity dispersion of an isolated stellar
disk}\label{sec:group}

Stars embedded in accretion disks are often considered in a ``test star''
regime \citep[e.g.,][]{Syer91}, when each star co-rotates with the accretion
disk. The star's velocity is then nearly equal to the local Keplerian circular
velocity.  When number of embedded stars, $N_*\gg 1$, two-body interactions
between stars will increase their local velocity dispersion, $\sigma$,
potentially leading to some interesting consequences.

Since the disk velocity field has the radial Keplerian shear, it is
the radial velocity dispersion of stars that will grow the fastest.
However, when the anisotropy $\sigma_r/\sigma_z$ becomes larger than
3, the buckling instability will develop and the stellar velocity
dispersion will become more isotropic \citep*{Kulsrud71,
Shlosman89}. Thus we assume an isotropic velocity dispersion here for
simplicity. The velocity dispersion of stars grows due to {\it N}-body
interactions at the rate
\begin{equation} 
\frac{d \sigma}{d t} \sim \frac{4 \pi G^2 M_* \rho_*}{\sigma^2}
\ln \Lambda_*
\label{dsdt}
\end{equation}
where $\Lambda_* \sim H_* \sigma^2/GM_*$ is the Coulomb logarithm for
stellar collisions; $H_*$ is the stellar disk height scale, which in
general may be different from the gas disk height scale $H$. 

The growth of velocity dispersions opposed by the dynamical friction force
acting between the stars and the gas.  Consider a star moving inside the disk
with a relative velocity $v_{\rm rel}$ with respect to the local Keplerian
velocity, $v_{\rm K}$.  \cite{Artymowicz94} shows that the angular momentum
and energy flow between the disk and the star (a small disk perturber in the
case of proto-planetary disks), calculated explicitly, coincides within a
factor of few with the hydrodynamical Bondi-Hoyle drag acting on the star
during its passage through the disk. The acceleration experienced by the star
is thus
\begin{eqnarray}
\vec{a_{\rm d}} = - 4 \pi G^2 M_* \rho C_{\rm d} \frac{\vec{v}_{\rm rel}}{g^4
(c_{\rm s}^2 + v_{\rm rel}^2)^{3/2}} \;,
\label{fd}
\end{eqnarray}
where $g = \min(1,v_{\rm c}/v_{\rm rel})$ and $v_{\rm c} = C_{\rm d}^{1/4}
v_{\rm esc,*}$ is of the order of the escape velocity from the star, $v_{\rm
esc,*}$. In perturbative analytical approaches, such as dynamical friction,
$C_{\rm d} \simgt 1$ is the Coulomb logarithm, $\ln \Lambda$, where $\Lambda$
is the ratio of the disk height scale $H$ to Bondi (or accretion) radius
\citep[e.g.,][]{Ostriker83}.  However, in many circumstances the Bondi-Hoyle
formula for accretion rate onto the star produces supper-Eddington values. The
drag force (e.g. $C_{\rm d}$) should then be reduced to account for the
radiation pressure force. One finds that in the disk geometry the largest
contribution to the star-gas friction comes from distances $\sim H$ from the
star. The exact value of $C_{\rm d}$ depends on disk opacity and the 3-D
velocity of the star, but estimates suggest that $C_{\rm d}$ is not much
smaller than unity in this case.

Note that when the relative velocity is high, the drag force is just
the hydrodynamical drag, $a_{\rm d} \propto \pi R_*^2 \rho v_{\rm rel}^2$,
where $R_*$ is the stellar radius. For $c_{\rm s} < v_{\rm rel} < v_{\rm esc,
*}$, the classical \cite{Chandrasekhar43}'s dynamical friction formula
is recovered, with $a_{\rm d} \propto v_{\rm rel}^{-2}$. Finally, if the relative
velocity is smaller than $c_{\rm s}$, we have $a_{\rm d} \sim \dot M_{\rm B}
v_{\rm rel}$, which is about equal to the momentum flux accreted by
the star ($\dot M_{\rm B}$ is the Bondi accretion rate).

While $v_{\rm rel}$ is not too large, i.e., $g=1$, the evolution of
the stellar velocity dispersion is approximately given by
\begin{equation} 
\frac{d \sigma}{d t} \sim 4 \pi G^2 M_* \left[ \frac{\rho_* \ln
\Lambda_*}{\sigma^2} - \frac{\rho C_{\rm d} \sigma}{(c_{\rm s}^2 +
\sigma^2)^{3/2}}\right]\;
\label{dsnet}
\end{equation}
As long as $\rho_* \ln \Lambda_* < \rho C_{\rm d}$, the star-gas drag will
be able to keep the stars on local circular Keplerian orbits in the
sense that $\sigma \ll c_{\rm s}$, the gas sound speed, thus the stars
indeed behave as test particles. However, when $\rho_* \ln \Lambda_* >
\rho C_{\rm d}$, the stellar velocity dispersion will evolve mainly under
the influence of {\it N}-body collisions, and it will run away.

It may appear that the last fact suggests a natural mechanism for
stopping the very efficient (see \S \ref{sec:discussion}) accretion of
gas onto the embedded (proto-) stars. As the stellar velocity
dispersion grows much larger than the gas sound speed, the stars will
be no longer embedded in accretion disks as they would spend most of
their orbits outside the main body of the accretion disk. In addition,
even when the stars are crossing the disk, the relatively high value
of $v_{\rm rel}$ means that the accretion rate onto stars will be
strongly reduced. However, the effect is important only when $\rho_* >
\rho$ (assuming $\ln \Lambda_* \sim C_{\rm d}$), that is when the
stellar density is already larger than the gas density. Therefore,
before this effect may become important, about a half of the initial
gas mass should already be consumed by the stars. The accretion onto
the stars is curbed by the {\it N}-body dispersion effects too late,
when the disk is already half eaten by the stars.

Now, coming back to the \sgra\ case, we can estimate the expected $H_*
\sim R \sigma/v_{\rm K}$.  The relaxation time, defined as the time
needed for the stellar disk to thicken to height $H_*$ can be found
from equation \ref{dsdt}:
\begin{equation} 
\frac{t_{\rm rel}}{t_{\rm dyn}} \sim \left[\frac{H_*}{R}\right]^4
\frac{M_{\rm BH}^2 }{4 M_{\rm d*} M_* \ln \Lambda_*}\;,
\label{trsph}
\end{equation}
where $M_{\rm d*}$ is $\pi R^2 H_* \rho_*$, the mass of the stellar disk.
Equation
\ref{trsph} yields
\begin{equation} 
\frac{t_{\rm rel}}{t_{\rm dyn}} \sim \; 2500 
\;\frac{10^4 \msun}{M_{\rm d*}} \;
\frac{10\msun}{M_*}\;\left[\frac{H_*/R}{0.1}\right]^4 \ln \Lambda_*^{-1}\;.
\label{trthin}
\end{equation}
With $t_{\rm dyn} \sim 300$ years, we have $t_{\rm rel} \sim 10^6$
years. Hence the geometrical thickness of the rings, and the ratio of
velocity dispersion to the local Keplerian velocity, $\sigma/v_{\rm K}$, are
expected to be of order $0.1$ for the two young stellar rings in the
GC. The individual stellar velocities should thus be still close to
the local {\em circular} Keplerian values if the origin of the stars
is in the gaseous disk.

\cite{Levin03} estimate the geometrical thickness of the inner stellar
disk in \sgra\ to be of order $H_*/R=0.1$. This ratio is however
larger but is not quantified for the outer disk found by
\cite{Genzel03}. From their figure 15 we estimate that $H_*/R\sim 0.3$
for the outer, counter-rotating, disk.

One may try to invert equation \ref{trthin} to constrain the initial
stellar mass of the disks in the GC by using the observed velocity
dispersions \citep{Genzel03} in the rings. Unfortunately the limits
are not very stringent due to the strong dependence of $t_{\rm rel}$
on $H/R$.  A disk mass as high as $M_{\rm d}\sim 3 \times 10^5\msun$
could still be consistent with the observations for the inner stellar
ring. Interestingly, for the outer stellar ring, the velocity
dispersion is too high to be explained by the {\it N}-body effects unless
the ring mass is unrealistically high.

\section{Destruction of stellar rings by orbital precession: the
maximum disk mass.}\label{sec:max}

\cite{Genzel03} find that most of the young innermost stars lie in one of {\em
two} stellar rings. There is no noticeable difference in the estimated age of
the two groups of stars. The rings are bound to interact gravitationally with
one another, and this could lead to observable disk distortions.

In particular, stellar orbits precess around the axis of symmetry in an
axisymmetric potential \citep[e.g., \S 3.2 in][]{Binney87}.  We represented
one of the disks by the Kuzmin potential
\begin{equation}
\Phi_K(R,z) = - \frac{G M_d}{\sqrt{R^2 + (a + |z|)^2}}\;,
\label{kuzmin}
\end{equation}
where $a$ is the disk radius, $R$ is the radius in the cylindrical
coordinates and $z$ is the perpendicular distance from the disk. We
then numerically integrated stellar orbits, starting from nearly
circular Keplerian orbits unperturbed by the disk presence. The orbits
remain approximately circular, and conserve the inclination angle $i$
between the orbital plane and the disk plane (because the
$z-$component of the angular momentum is rigorously conserved in the
axisymmetric potential). The stellar orbital plane precesses with
respect to the disk at a rate
\begin{equation}
\dot{\phi} = C_{\rm p} q P^{-1} \cos i\;,
\label{phip}
\end{equation}
where $C_{\rm p}$ is a constant (for a given orbit and given geometry)
of order unity.  Angle $\phi$ here is the azimuthal angle of the lines
of the nodes for the orbit in cylindrical coordinates used to define
the Kuzmin potential. The precession rate scaling (equation
\ref{phip}) is natural since for small $q$ the effect is linear in $q$
as can be seen for orbits nearly co-planar with the disk (when $i
\approx 0^\circ$); for $i=90^\circ$ there should be no plane
precession due to symmetry.

The value of $C_{\rm p}$ depends on the value of $a$ with respect to the
radius of the nearly circular stellar orbit; for $a$ of order the
radius, $C_{\rm p} \sim 1 $. Setting $i = 74^\circ$ as appropriate for the
two GC stellar rings \citep{Genzel03}, we obtain
\begin{equation}
\Delta \phi \simeq \; C_{\rm p} \; \frac{q}{0.003} \frac{t}{10^3 P}\;.
\label{dphi}
\end{equation}

The important point to note is that nearly circular orbits of stars at
different radii from the SMBH will precess by different amounts
$\Delta \phi$. Therefore such a precession leads to a warping of the
stellar disk. After a time long enough to yield $\Delta \phi \simgt 1$
somewhere in the disk, the initially flat stellar disk will be
disfigured and will not be recognizable as a disk at all by an
observer.

An approximate upper limit on the {\em time-average} mass of each of the two
stellar rings in \sgra\ can be set. Clearly the exact value of this limit
should be obtained numerically with {\it N}-body experiments and comparison
with the quality ($\chi^2$) of the fits to the two observed planes
\citep{Levin03, Genzel03}. Such a study is underway. Due to an observational
uncertainty in the radial dimensions of the rings' inner and outer radii, and
theoretical uncertainty in the distribution of gaseous mass (i.e. $\Sigma(R)$)
in the accretion disk, it is possible to reduce $C_{\rm p}$ from its maximum
value for some values of parameters. Nevertheless, a rather robust value for
the upper mass of the disks appears to be
\begin{equation}
\max{M_{\rm d}}\approx 10^5 \msun\;.
\label{mdmax}
\end{equation}

\section{Rotating the accretion disk midplane: do stars remain
embedded?}\label{sec:two}

The accretion disk midplane orientation can in principle change as a
result of a new mass deposition coming with a different orientation of
the angular momentum vector. In such a rotation, would the newly born
stars remain embedded into the disk and follow its rotation or would
they stay behind in the ``old'' accretion disk midplane due to their
large inertia?  The answer to this question is important for AGN disks
in general as embedded stars can seriously influence the accretion
process \citep[e.g.,][]{GoodmanTan04, Nayakshin04}.

By order of magnitude, one can estimate the time needed to turn the
accretion disk plane around on a significant angle to be
\begin{equation}
t_{\rm rot} \sim \frac{M_{\rm d}}{\dot{M}_{\rm c}} \sim t_{\rm visc}
\frac{\dot{M}}{\dot{M}_{\rm c}}\;,
\label{trot}
\end{equation}
where $\dot{M}_{\rm c}$ is the mass condensation rate onto the
accretion disk.  If the accretion and condensation processes are in an
approximate steady state, $\dot{M}_{\rm c} = \dot{M}$, $t_{\rm rot}
\sim t_{\rm visc}$. The latter is
\begin{equation}
t_{\rm visc} \sim \frac{R}{v_{\rm K}} \alpha^{-1} \;
\left(\frac{R}{H}\right)^2\;,
\end{equation}
and can be fairly long. Thus in general the disk plane orientation
changes rather slowly. 

\subsection{Forces keeping the stars embedded}\label{sec:forces}

Two forces mediating interaction between a star and a gaseous disk are
the gravity of the disk as a whole, and the friction acting on a star
{\em moving} inside the disk at a certain velocity with respect to the
local circular Keplerian speed. If stars lag behind the rotating disk
plane, the characteristic relative velocity at which the star and the
gas would be separated is $\sim v_{\rm K}/t_{\rm rot}$ and is very
small compared to the sound speed in the gas (if $t_{\rm rot}\sim
t_{\rm visc}$).  At small relative velocities the dynamical friction
of a star ``leaving'' the gas disk is very small too (see equation
\ref{fd}), and a simple estimate shows that the dynamical friction
force can be safely neglected in what follows below.

Therefore the binding force to consider is the direct gravitational
attraction between the disk and the star. Near the disk midplane, the
infinite plane approximation can be used for the disk gravity. The
gravitational attraction force of the gaseous disk for a star that
left the disk midplane (i.e., the star-disk midplane separation
$|z|\simgt H$) is
\begin{equation}
a_{\rm pl} = 2 \pi G \Sigma\;, 
\label{apl}
\end{equation}
where $\Sigma$ is the local disk surface density. Comparing this
acceleration with the centrifugal acceleration of the star moving in a
circular Keplerian orbit around the central black hole, $a_{\rm c} =
v_{\rm K}^2/R$,
\begin{equation}
\frac{a_{\rm pl}}{a_{\rm c}} = \frac{2 M_{\rm d}}{M_{\rm BH}}\;.
\label{atoac}
\end{equation}

\subsection{Critical rotation time}\label{sec:test}

Suppose that the accretion disk midplane turns at a rate given by the
time scale $t_{\rm rot}$. Define a critical rotation time scale,
$t_{\rm rc}$, such that for disk plane changes occurring on time
scales shorter than $t_{\rm rc}$, the stars are dislodged from the gas
disk. For $t_{\rm rot} > t_{\rm rc}$, on the contrary, the stars
remain bound to the disk. Clearly, we get the critical time scale when
$a_{\rm pl} = a_{\rm rot} \equiv v_K/t_{\rm rot}$, where $a_{\rm rot}$
is the ``rotation acceleration'' of the turning disk midplane. We
obtain for the critical rotation time
\begin{equation}
t_{\rm rc} = \frac{M_{\rm BH}}{2 M_{\rm d}} t_{\rm dyn}\;.
\label{trc}
\end{equation}
Figure \ref{fig:disk3e6b} shows the critical rotation time scales
(dotted curves) along with other important time scales for the
standard accretion disk model with same parameters as used for Fig.
\ref{fig:disk3e6a} and for a 10 Solar mass star. The thick line curves
are for $\mdot=0.03$, whereas the thin curves are for $\mdot=1$. The
accretion and migration time scales will be discussed in \S
\ref{sec:accretion} below.

Note that $t_{\rm rc}$ is longer than $t_{\rm dyn}$ but is much
shorter than $t_{\rm visc}$. This implies that if accretion disk plane
changes occur on a viscous time scale, the stars would remain bound to
the disk. Only very fast plane changes could dislodge the stars from
the disk midplane.

\subsection{The case of \sgra}

We have just shown that it is fairly difficult to separate the stars
and the accretion disks in slow disk plane rotations or
deformations. For the \sgra\ case, this implies that either (i) there
were two separate accretion disk creation events that created the two
differently oriented rings; or (ii) the accretion disk itself was
extremely warped so that its inner part was oriented almost at the
right angle with respect to the outer disk part.

\section{Accretion onto embedded stars}\label{sec:accretion}

\begin{figure}
\centerline{\psfig{file=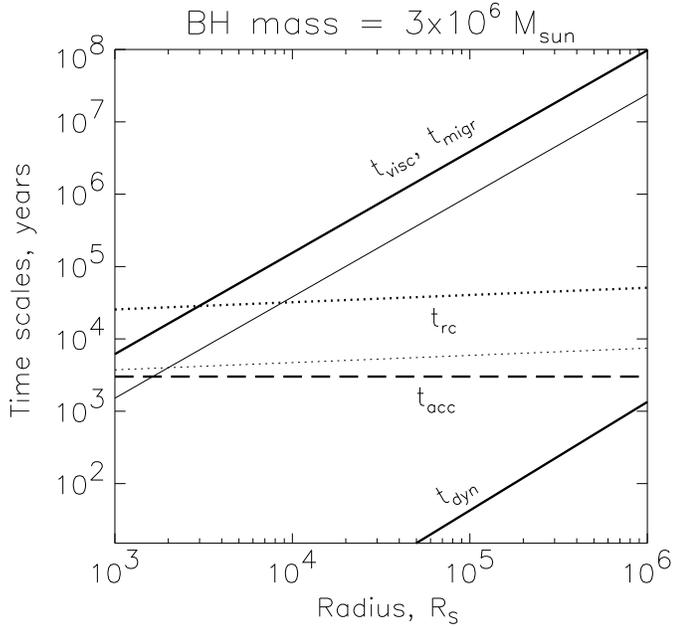,width=.49\textwidth}}
\caption{Time scales for a 10 Solar mass star embedded in the standard
accretion disk with parameters as in Fig. \ref{fig:disk3e6a}. As
before, thick curves correspond to $\mdot=0.03$, whereas thin ones are
for $\mdot=1$. The solid lines show the viscous and the dynamical time
scales for the disk, as labelled in the figure. The star is massive
enough to open up a gap and hence migrates inward on the viscous
timescale ($t_{\rm migr} = t_{\rm visc}$).  The dashed and dotted
lines are the accretion and the critical rotation time scales,
respectively. For chosen parameters, the former one is independent of
$\mdot$ (see text in \S \ref{sec:accretion} for detail).}
\label{fig:disk3e6b}
\end{figure}

The Hill's radius $R_H$,
\begin{equation}
R_H = \left[\frac{M_*}{3 M_{\rm BH}}\right]^{1/3}\,R\;,
\label{rh}
\end{equation} 
defines the sphere around the star where the dynamics of gas is
dominated by the star rather than the SMBH.  The accretion of gas onto
a star is believed to be similar to the growth of terrestrial planets
in a planetesimal disk \citep{Lissauer87,Bate03}. For $R_H >
H$, gas accretion onto the star is quasi two-dimensional. The
accretion rate is determined by the rate at which differential
rotation brings the matter into the Hill's sphere,
\begin{equation}
\dot{M}_* = \dot{M}_H \sim 4 \pi R_H H\, \rho \,v_H \sim 4 \pi R_H^2
\rho c_s \;,
\label{mhill}
\end{equation}
where $\rho = \Sigma/2H$ is the mean disk density. We used the fact
that the characteristic gas velocity (relative to the star) at the
Hill's distance from the star, $v_H$, is $v_H = R_H |d \Omega/d\ln R|
\sim c_s (R_H/H)$ since the angular velocity for Keplerian rotation is
$\Omega = c_s/H$.  Equation \ref{mhill} is valid as long as $R_H > H$
since in the opposite case the gas thermal velocity becomes important
and the accretion would proceed at the Bondi accretion rate
\citep[$\dot{M}_{\rm B}$; e.g.,][]{Syer91}.  Of course $\dot{M}_*$
cannot exceed $\dot{M}_{\rm *,Edd} \simeq 10^{-3} r_* \simeq 10^{-3}
\msun/\hbox{year}\; m_*^{1/2}$, the Eddington accretion rate onto the
star \footnote{We assumed that $r_* = (R_*/\rsun) \approx
(M_*/\msun)^{1/2}$. Note that equation 10 in \cite{Nayakshin04} contains a
typo.}. We thus estimate
\begin{equation}
\dot{M}_* = \hbox{min}\;\left[\dot{M}_H, \dot{M}_{\rm B}, \dot{M}_{\rm
*,Edd}\right]\;.
\label{mstar}
\end{equation}
One can then define the accretion time scale for a star embedded into
a disk:
\begin{equation}
t_{\rm acc} \equiv \frac{M_*}{\dot{M}_*}\;.
\label{tstar}
\end{equation}
Figure \ref{fig:disk3e6b} shows the accretion time scale (dashed line)
for a 10 $\msun$ embedded star. Although we considered two values for
the accretion rate onto the SMBH, $\mdot=0.03$ and $\mdot=1$, as in
Figure \ref{fig:disk3e6a}, $t_{\rm acc}$ turns out to be the same for
both of these because the accretion rate is close to the Eddington
value.

An important point to take from Figure \ref{fig:disk3e6b} is that
accretion onto embedded stars is able to double the stellar mass in
few thousand years. Therefore, growing stars as massive as 100 $\msun$
in a million years in a disk with gas mass $M_d \simgt 10^4 \msun$
appears to be no problem at all. [One potential uncertainty here is
the reduction in the accretion rate onto the embedded stars once these
stars are massive enough to clear out a radial gap in the accretion
disk. Results of \cite{Bate03}, Figure 9, show that this reduction
can be very large. However, in the case of an AGN disk with {\em many}
embedded stars, the dynamics of the star-gas interaction is surely
going to be different from the case of a ``test'' star or planet. The
accretion disk will then be divided onto many rings between stars on
nearly circular radial orbits. If the orbits are close enough (number
of stars $N_*\gg 1$), then the gas in a ring will experience
alternating inward and outward pushes from the two stars closest to it
and hence the radial gap can in fact be closed, enabling an unhindered
accretion. The issue deserves a future study.]

We also estimated the radial migration time scale, $t_{\rm migr}$,
using the prescription for the radial migration velocity based on the
numerical calculations of \cite{Bate03}. For the parameters
chosen, the 10 $\msun$ stars, and any stars more massive than that,
open up a gap in the accretion disk and their radial migration is
identical with the viscous flow of matter in the gas disk. Thus
$t_{\rm migr} = t_{\rm visc}$ (two solid curves in Figure
\ref{fig:disk3e6b}). The migration time scale is very long, indicating
that stars will remain pretty much where they were born in accretion
disks with parameters close to that of the standard disk for \sgra. In
fact, a more realistic self-gravitating disk would not change this
conclusion significantly since the migration time scale only gets
longer when the midplane disk density decreases as a result of disk
swelling due to gravitationally induced turbulence.

\section{Growth of ``mini star clusters'' and intermediate mass black
holes in accretion disks}\label{sec:cluster}

\cite{GoodmanTan04} have recently suggested that it is possible to
grow supermassive stars in AGN accretion disks. The maximum mass of a
star in this case is the gas disk mass in a ring with width of order
the Hill's radius of the star, $R_H = (M_*/3 M_{\rm BH})^{1/3}$. This
is the ``isolation'' mass, $M_i\simeq M_d^{3/2} M_{\rm BH}^{-1/2}$,
\begin{equation}
M_i \approx 550 \msun \;
\left[\frac{M_d}{10^4 \msun}\right]^{3/2} \left[\frac{3\times 10^6
\msun}{M_{\rm BH}}\right]^{1/2}\;.
\label{miso}
\end{equation}
The stability of super-massive stars is briefly summarized in \S 2 of
\citep{GoodmanTan04}. The supermassive star could collapse directly
into a black hole if the star is more massive than 300 $\msun$ \citep{Fryer01}.

However, the Hill's accretion rate estimate assumes that all of the
disk mass delivered by the differential rotation into the Hill's
radius about the star is accreted onto the star. Even without the gap
presence, this is not obvious because the gas still has to loose most
of its angular momentum before it will reach the stellar surface
\citep[e.g.,][]{Milosavljevic04}. Furthermore, quite frequently the
accretion rate onto the star estimated in this way exceeds the
Eddington accretion rate onto the star (as is the case for Figure
2). \cite{Milosavljevic04} have shown that the fringes of the
small scale disk around the embedded stars themselves become
self-gravitating and may therefore also form stars or planets. It is
thus possible to grow in situ star clusters. The maximum total mass of
such a cluster should be close to the isolation mass.

A qualitative confirmation of these ideas can be found in numerical
simulations of a related physical problem by \cite{Tanga04}. These
authors simulate the growth and clustering of planetesimals in a
proto-stellar disk. They find a hierarchical growth of clusters of
particles and find that these ``clusters'' are intrinsically stable
structures. This is likely because of the abundant supply of gas into
the Hill's sphere: there is always a plenty of gas to interact with
the particles (gravitationally bound objects in AGN case) that
continue to get more and more bound to the central object in the
mini-cluster.

This mechanism of intermediate mass black hole (IMBH) and bound to it
star cluster may be relevant to the observations of the IRS13 cluster
near \sgra \citep{Maillard04}. The ``dark'' mass in the IRS13 is
estimated to be $\simgt 10^3 \msun$. Equation \ref{miso} shows that an
initial mass of the disk of order several times $10^4 \msun$ would
have been sufficient to grow ``in situ'' an object massive enough to
become the IRS13 cluster.

\section{Discussion}\label{sec:discussion}

We have considered here the formation of massive stars in a
self-gravitating accretion disk for conditions appropriate for the
central $\sim 0.2$ pc of our Galaxy.  Formation of an accretion disk
(instead of a narrow ring) would be a likely outcome of a cooling
instability for a hot gas since the gas would realistically have a
broad range of the angular momentum values. Additionally, a cloud with
an initial size of a parsec or larger, tidally disrupted and shocked,
should settle in a disk of a size comparable to its initial
radius. There is of course a direct observational test which would
distinguish between the accretion disk versus the compact infalling
cloud idea of \cite{Sanders98} -- one simply has to establish whether
the orbits of the young stars in the two stellar rings are nearly
circular or they are strongly eccentric. As we showed in \S
\ref{sec:group}, stars born in an accretion disk in \sgra\ would still
retain their nearly circular orbits.

The standard theory of gravitational instability for a thin disk
\citep[]{Toomre64,Paczynski78} predicts that the minimum mass of the
gas in the disk that would make it gravitationally unstable for the
parameters appropriate to our Galactic Center is $\sim 10^4 \msun$ (\S
\ref{sec:selfgrav}). It would be interesting to compare the predicted
stellar mass resulting from star formation in such an accretion disk
with the current stellar content of the rings. Unfortunately
theoretical uncertainties for the efficiency of star formation in
self-gravitating disks are too large.  \cite*{Shlosman90} have shown
that if the cooling time of a self-gravitating disk is shorter than
$t_{\rm dyn} = \Omega^{-1}$, then the disk will fragment and form
stars and/or planets. For longer cooling times, it was argued that the
disk does not fragment \citep{Shlosman90}. Numerical simulations with
a constant cooling time by \cite{Gammie01} confirmed this, and have
shown that the disk settles into a stable state where the cooling is
offset by the energy input generated by gravitational instabilities
\citep[see also][]{Paczynski78}. Yet for disk temperatures of order
$10^3$ Kelvin, the opacity is strongly dependent on the temperature.
\cite{Johnson03} showed that in the non-linear stage of the
instabilities, the local cooling time may be orders of magnitude
smaller than that found in the unperturbed disk model. However AGN
disks are usually hotter than this and hence the non-linear effect
should be weaker.

Nevertheless, we believe that \sgra\ accretion disk was likely
consumed almost entirely in the star formation episode rather than has
been accreted by the SMBH. There is no doubt about star formation
here: there are dozens of brightest and quite massive stars in each of
the stellar rings with 3-D velocity measurements. There are additional
numbers of dimmer stars that have only 2-D velocities measured but are
strongly suspected of belonging to these same rings \cite[ \S
3.7]{Genzel03}. The accretion time scale on the embedded stars is very
short (\S \ref{sec:accretion} and Figure \ref{fig:disk3e6b}) compared
to the disk viscous time scale. We have also shown in this paper, that
neither disk plane rotations, warps, or the stellar {\it N}-body
scattering \citep[unless there is already more mass in the stars than
in the gas, see also][] {CN04} can ``shake'' the stars off the disk
midplane. In addition, each massive star opens up a radial gap in the
accretion disk around it. These stars would not let {\em the standard}
accretion disk to transfer the gas into the SMBH simply because they
are in the way of the gas flow.

While the standard accretion disk equations are not applicable to the
region where the disk becomes self-gravitating, the stellar accretion
time scale is shorter than the viscous time by 3-4 orders of
magnitude. We experimented with a prescription for the accretion disk
equations which introduces turbulent energy and pressure in addition
to the thermal ones to keep the disk marginally stable (i.e., $Q\simgt
1$), and found that the turbulent energy content should be
unrealistically high to reduce the accretion rate onto the stars
sufficiently.

We have also found (\S \ref{sec:group}) that the geometrical thickness
of the inner $2-4''$ stellar ring, and the stellar velocity
dispersion, could be explained by {\it N}-body scatterings between the
members of the same ring if the initial stellar mass was as high as
$10^4 \msun$. However, the outer $4-7''$ projected distance ring
\citep{Genzel03} is too thick to result from the internal
scatterings. We believe that a better explanation is stellar ring
warping due to a non-spherical gravitational potential (\S
\ref{sec:two}), e.g. due to the presence of the inner ring. Estimating
the rate at which the rings get distorted, we tentatively set an upper
limit on the {\em time-average} total mass of each of the gas-star
disks (rings) at around $10^5 \msun$. Future numerical {\it N}-body
modeling and direct comparison to stellar orbits may tighten this
limit.

We hope that future observations of the stellar mass content in the
two rings in \sgra, and also observations of the inner Galaxy ISM
budget, could be used to constrain the initial mass of the gaseous
accretion disk, and its further fate. As we have shown, the gaseous
disk mass should have been in the range of $(1-10) \times 10^4 \msun$,
10 to 100 times higher than the present day mass in the observed
stellar rings \citep{Genzel03}. If most of the disk gas was used to
make stars, then more of these stars and/or their remnants should be
found in the future in the inner $\sim 0.2$ pc of the Galaxy. In
addition, one may look for evidence of a hot high metallicity bubble
in the inner 1 kpc of the Galaxy produced by stellar winds and
supernova explosions.

If instead the gas was mostly accreted by \sgra, then there should be
evidence of a past quasar phase. The required accretion rate, $\sim
(10^4-10^5) \msun/10^6$~year $ = 10^{-2}$ to $10^{-1}
\msun$~year$^{-1}$ is comparable with the Eddington accretion rate for
\sgra, $\medd \sim 0.03 \msun$ year$^{-1}$. This would have to be a
very rare event in \sgra\ recent life since the other nearby galaxies
either have no AGN or have very weak ones with X-ray luminosity
usually smaller than $L_X\simlt 10^{40}$ erg sec$^{-1}$
\cite[e.g.,][]{Zang01}. A hot buoyant radio bubble would most likely
be present in the Milky Way halo, as accretion onto the SMBH is widely
believed to go hand in hand with superluminous jet outflows.

\section{Conclusions}\label{sec:conclusions}

Our main results are as following:

1. The minimum mass of each of the disks needed to form the observed
   young stars by self-gravity is around $10^4\msun$.

2. The observed stellar velocity dispersions in the outer ring is too
   large to result from {\it N}-body interactions between stars belonging
   to the same ring. The orbital precession of stars caused by the
   potential of the other disk can explain the observed disk thickness
   and velocity dispersion if the time average stellar and gaseous
   mass in the inner disk is in the range $(3-10) \times 10^4 \msun$.

3. Few million years ago, \sgra\ had a good chance to become a very
bright AGN with the bolometric luminosity $L\sim 10^{44} - 10^{45}$
erg/sec, but was robbed of most of its gaseous fuel by nuclear star
formation in a self-gravitating accretion disk. Nevertheless, even if
only a few percent of the available disk fuel was captured by \sgra,
the SMBH in our Galactic Center was as bright as $L\sim 10^{42} -
10^{43}$ erg/sec.

We have also shown that capture of stars from the ``old'' relaxed
isotropic \sgra\ star cluster \citep[the cusp; see][]{Genzel03} is
inefficient unless the gaseous disk mass were as high as $10^5
\msun$. In addition, the role of possible accretion disk midplane
changes was estimated. It was found that the embedded stars inertia
would have been efficient in taking the stars out of the body of the
disks only if the disk plane changes its orientation on time scales
much shorter than the disk viscous time.

\acknowledgement The authors acknowledge useful discussions with
Eugene Churazov, Walter Dehnen, Reinhard Genzel, Marat Gilfanov, Peter
Predehl and Rashid Sunyaev.

\bibliography{aamnem99,jcuadra}

\begin{thebibliography}{47}
\expandafter\ifx\csname natexlab\endcsname\relax\def\natexlab#1{#1}\fi

\bibitem[{{Artymowicz}(1994)}]{Artymowicz94}
{Artymowicz}, P. 1994, \apj, 423, 581

\bibitem[{{Artymowicz} {et~al.}(1993){Artymowicz}, {Lin}, \&
  {Wampler}}]{Artymowicz93}
{Artymowicz}, P., {Lin}, D.~N.~C., \& {Wampler}, E.~J. 1993, \apj, 409, 592

\bibitem[{{Bate} {et~al.}(2003){Bate}, {Lubow}, {Ogilvie}, \&
  {Miller}}]{Bate03}
{Bate}, M.~R., {Lubow}, S.~H., {Ogilvie}, G.~I., \& {Miller}, K.~A. 2003,
  \mnras, 341, 213

\bibitem[{{Binney} \& {Tremaine}(1987)}]{Binney87}
{Binney}, J. \& {Tremaine}, S. 1987, {Galactic dynamics} (Princeton University
  Press)

\bibitem[{{Chandrasekhar}(1943)}]{Chandrasekhar43}
{Chandrasekhar}, S. 1943, \apj, 97, 255

\bibitem[{{Collin} \& {Zahn}(1999)}]{Collin99}
{Collin}, S. \& {Zahn}, J. 1999, \aap, 344, 433

\bibitem[{{Cuadra} \& {Nayakshin}(2004)}]{CN04}
{Cuadra}, J. \& {Nayakshin}, S. 2004, in Growing Black Holes: Accretion in a
  Cosmological Context, ed. A. Merloni, S. Nayakshin \& R. Sunyaev (in press)

\bibitem[{{Fryer} {et~al.}(2001){Fryer}, {Woosley}, \& {Heger}}]{Fryer01}
{Fryer}, C.~L., {Woosley}, S.~E., \& {Heger}, A. 2001, \apj, 550, 372

\bibitem[{{Gammie}(2001)}]{Gammie01}
{Gammie}, C.~F. 2001, \apj, 553, 174

\bibitem[{{Genzel} {et~al.}(2003){Genzel}, {Sch{\" o}del}, {Ott}, {Eisenhauer},
  {Hofmann}, {Lehnert}, {Eckart}, {Alexander}, {Sternberg}, {Lenzen}, {Cl{\'
  e}net}, {Lacombe}, {Rouan}, {Renzini}, \& {Tacconi-Garman}}]{Genzel03}
{Genzel}, R., {Sch{\" o}del}, R., {Ott}, T., {et~al.} 2003, \apj, 594, 812

\bibitem[{{Gerhard}(2001)}]{Gerhard01}
{Gerhard}, O. 2001, \apj, 546, L39

\bibitem[{{Ghez} {et~al.}(2003){Ghez}, {Duch{\^ e}ne}, {Matthews}, {Hornstein},
  {Tanner}, {Larkin}, {Morris}, {Becklin}, {Salim}, {Kremenek}, {Thompson},
  {Soifer}, {Neugebauer}, \& {McLean}}]{Ghez03b}
{Ghez}, A.~M., {Duch{\^ e}ne}, G., {Matthews}, K., {et~al.} 2003, \apj, 586,
  L127

\bibitem[{{Goodman}(2003)}]{Goodman03}
{Goodman}, J. 2003, \mnras, 339, 937

\bibitem[{{Goodman} \& {Tan}(2004)}]{GoodmanTan04}
{Goodman}, J. \& {Tan}, J.~C. 2004, \apj, 608, 108

\bibitem[{{Hansen} \& {Milosavljevi{\' c}}(2003)}]{Hansen03}
{Hansen}, B.~M.~S. \& {Milosavljevi{\' c}}, M. 2003, \apj, 593, L77

\bibitem[{{Johnson} \& {Gammie}(2003)}]{Johnson03}
{Johnson}, B.~M. \& {Gammie}, C.~F. 2003, \apj, 597, 131

\bibitem[{{Kim} {et~al.}(2004){Kim}, {Figer}, \& {Morris}}]{Kim04}
{Kim}, S.~S., {Figer}, D.~F., \& {Morris}, M. 2004, \apj, 607, L123

\bibitem[{{Kim} \& {Morris}(2003)}]{Kim03}
{Kim}, S.~S. \& {Morris}, M. 2003, \apj, 597, 312

\bibitem[{{Kolykhalov} \& {Sunyaev}(1980)}]{Kolykhalov80}
{Kolykhalov}, P.~I. \& {Sunyaev}, R.~A. 1980, Soviet Astron. Lett., 6, 357

\bibitem[{{Koyama} {et~al.}(1996){Koyama}, {Maeda}, {Sonobe}, {Takeshima},
  {Tanaka}, \& {Yamauchi}}]{Koyama96}
{Koyama}, K., {Maeda}, Y., {Sonobe}, T., {et~al.} 1996, PASJ, 48, 249

\bibitem[{{Krabbe} {et~al.}(1995){Krabbe}, {Genzel}, {Eckart}, {Najarro},
  {Lutz}, {Cameron}, {Kroker}, {Tacconi-Garman}, {Thatte}, {Weitzel},
  {Drapatz}, {Geballe}, {Sternberg}, \& {Kudritzki}}]{Krabbe95}
{Krabbe}, A., {Genzel}, R., {Eckart}, A., {et~al.} 1995, \apj, 447, L95

\bibitem[{{Kulsrud} {et~al.}(1971){Kulsrud}, {Mark}, \& {Caruso}}]{Kulsrud71}
{Kulsrud}, R.~M., {Mark}, J.~W.~K., \& {Caruso}, A. 1971, Ap\&SS, 14, 52

\bibitem[{{Levin} \& {Beloborodov}(2003)}]{Levin03}
{Levin}, Y. \& {Beloborodov}, A.~M. 2003, \apj, 590, L33

\bibitem[{{Lissauer}(1987)}]{Lissauer87}
{Lissauer}, J.~J. 1987, Icarus, 69, 249

\bibitem[{{Liszt}(2003)}]{Liszt03}
{Liszt}, H.~S. 2003, \aap, 408, 1009

\bibitem[{{Maillard} {et~al.}(2004){Maillard}, {Paumard}, {Stolovy}, \&
  {Rigaut}}]{Maillard04}
{Maillard}, J.~P., {Paumard}, T., {Stolovy}, S.~R., \& {Rigaut}, F. 2004, \aap,
  423, 155

\bibitem[{{McMillan} \& {Portegies Zwart}(2003)}]{McMillan03}
{McMillan}, S.~L.~W. \& {Portegies Zwart}, S.~F. 2003, \apj, 596, 314

\bibitem[{{Milosavljevi{\' c}} \& {Loeb}(2004)}]{Milosavljevic04}
{Milosavljevi{\' c}}, M. \& {Loeb}, A. 2004, \apj, 604, L45

\bibitem[{{Morris} {et~al.}(1999){Morris}, {Ghez}, \& {Becklin}}]{Morris99}
{Morris}, M., {Ghez}, A.~M., \& {Becklin}, E.~E. 1999, Advances in Space
  Research, 23, 959

\bibitem[{{Nayakshin}(2004)}]{Nayakshin04}
{Nayakshin}, S. 2004, \mnras, 352, 1028

\bibitem[{{Nayakshin} {et~al.}(2004){Nayakshin}, {Cuadra}, \&
  {Sunyaev}}]{NCS04}
{Nayakshin}, S., {Cuadra}, J., \& {Sunyaev}, R. 2004, \aap, 413, 173

\bibitem[{{Ostriker}(1983)}]{Ostriker83}
{Ostriker}, J.~P. 1983, \apj, 273, 99

\bibitem[{{Paczy\'nski}(1978)}]{Paczynski78}
{Paczy\'nski}, B. 1978, Acta Astron., 28, 91

\bibitem[{{Rees}(2002)}]{Rees02}
{Rees}, M.~J. 2002, in Lighthouses of the universe: the most luminous celestial
  objects and their use for cosmology, ed. M.Gilfanov, R. Sunyaev \& E.
  Churazov (Berlin: Springer), 345

\bibitem[{{Revnivtsev} {et~al.}(2004){Revnivtsev}, {Churazov}, {Sazonov},
  {Sunyaev}, {Lutovinov}, {Gilfanov}, {Vikhlinin}, {Shtykovsky}, \&
  {Pavlinsky}}]{Revnivtsev04}
{Revnivtsev}, M.~G., {Churazov}, E.~M., {Sazonov}, S.~Y., {et~al.} 2004, ArXiv
  Astrophysics e-prints, {\tt astro-ph/0408190}

\bibitem[{{Sanders}(1998)}]{Sanders98}
{Sanders}, R.~H. 1998, \mnras, 294, 35

\bibitem[{{Sch{\" o}del} {et~al.}(2002){Sch{\" o}del}, {Ott}, {Genzel},
  {Hofmann}, {Lehnert}, {Eckart}, {Mouawad}, {Alexander}, {Reid}, {Lenzen},
  {Hartung}, {Lacombe}, {Rouan}, {Gendron}, {Rousset}, {Lagrange}, {Brandner},
  {Ageorges}, {Lidman}, {Moorwood}, {Spyromilio}, {Hubin}, \&
  {Menten}}]{Schoedel02}
{Sch{\" o}del}, R., {Ott}, T., {Genzel}, R., {et~al.} 2002, \nat, 419, 694

\bibitem[{{Shakura} \& {Sunyaev}(1973)}]{Shakura73}
{Shakura}, N.~I. \& {Sunyaev}, R.~A. 1973, \aap, 24, 337

\bibitem[{{Shlosman} \& {Begelman}(1989)}]{Shlosman89}
{Shlosman}, I. \& {Begelman}, M.~C. 1989, \apj, 341, 685

\bibitem[{{Shlosman} {et~al.}(1990){Shlosman}, {Begelman}, \&
  {Frank}}]{Shlosman90}
{Shlosman}, I., {Begelman}, M.~C., \& {Frank}, J. 1990, \nat, 345, 679

\bibitem[{{Sunyaev} {et~al.}(1993){Sunyaev}, {Markevitch}, \&
  {Pavlinsky}}]{Sunyaev93}
{Sunyaev}, R.~A., {Markevitch}, M., \& {Pavlinsky}, M. 1993, \apj, 407, 606

\bibitem[{{Svensson} \& {Zdziarski}(1994)}]{Svensson94}
{Svensson}, R. \& {Zdziarski}, A.~A. 1994, \apj, 436, 599

\bibitem[{{Syer} {et~al.}(1991){Syer}, {Clarke}, \& {Rees}}]{Syer91}
{Syer}, D., {Clarke}, C.~J., \& {Rees}, M.~J. 1991, \mnras, 250, 505

\bibitem[{{Tanga} {et~al.}(2004){Tanga}, {Weidenschilling}, {Michel}, \&
  {Richardson}}]{Tanga04}
{Tanga}, P., {Weidenschilling}, S.~J., {Michel}, P., \& {Richardson}, D.~C.
  2004, ArXiv Astrophysics e-prints, {\tt astro-ph/0408441}

\bibitem[{{Toomre}(1964)}]{Toomre64}
{Toomre}, A. 1964, \apj, 139, 1217

\bibitem[{{Yu} \& {Tremaine}(2002)}]{Yu02}
{Yu}, Q. \& {Tremaine}, S. 2002, \mnras, 335, 965

\bibitem[{{Zang} \& {Meurs}(2001)}]{Zang01}
{Zang}, Z. \& {Meurs}, E.~J.~A. 2001, \apj, 556, 24

\end{thebibliography}
\bibliographystyle{aa}

\end{document}